\documentstyle[epsf,aps,prb,multicol]{revtex}
\begin{document}
\newcommand{\Pvec}{{\rm\bf P}}
\newcommand{\Evec}{{\rm\bf E}}
\newcommand{\eps}{\epsilon}  
\newcommand{\veps}{\varepsilon}  
\newcommand{\De}{$\Delta$}
\newcommand{\de}{$\delta$}
\newcommand{\mc}{\multicolumn}
\newcommand{\be}{\begin{eqnarray}}
\newcommand{\ee}{\end{eqnarray}}
\newcommand{\einf}{\varepsilon^\infty}
\newcommand{\ez}{\varepsilon^0}  

\draft \title{Effects of macroscopic polarization  in  
III-V nitride  multi-quantum-wells}
\author{Vincenzo Fiorentini}
\address{Istituto Nazionale per la Fisica della Materia 
-- Dipartimento di 
 Fisica, Universit\`a di Cagliari, Cagliari, Italy, and \\
 Walter Schottky Institut, Technische Universit\"a{}t
M\"u{}nchen, Garching, Germany}
\author{Fabio Bernardini}
\address{Istituto Nazionale per la Fisica della Materia 
-- Dipartimento di 
 Fisica, Universit\`a di Cagliari, Cagliari, Italy}
\author{Fabio Della Sala, Aldo Di Carlo, and Paolo Lugli}
\address{Istituto Nazionale per la Fisica della Materia 
-- Dipartimento  di Ingegneria Elettronica, 
Universit\`a di Roma ``Tor Vergata'', Roma, Italy}
\date{resub to PRB 28 dec 99} 
\maketitle

\begin{abstract}
Huge built-in electric fields have been  predicted to exist in 
wurtzite  III-V nitrides  thin films and multilayers.
Such fields originate from heterointerface discontinuities 
of the macroscopic bulk polarization of the nitrides.
Here we discuss the background theory, the role of spontaneous 
polarization in  this context, and  the   practical implications of
built-in polarization fields in nitride nanostructures. To support our
arguments, we present  detailed self-consistent tight-binding
simulations of typical nitride QW structures in which
polarization effects are dominant.  
\end{abstract}


\begin{multicols}{2}
\section{Introduction}

Spontaneous polarization has long been known to take place
in ferroelectrics. On the other hand, its existence in 
semiconductors with sufficiently low crystal symmetry (wurtzite, at
the very least)  has been  generally regarded as of purely theoretical
interest. Recently,  a series of first principles calculations
\cite{noi.piezo,noi.eps,noi.pol}  has reopened this issue for the 
 technologically relevant III-V nitride semiconductors, whose 
natural crystal structure is, in fact, wurtzite. Firstly,
\cite{noi.piezo} it was shown that the nitrides have a very large
spontaneous polarization, as well as  large piezoelectric
coupling constants.  Secondly, \cite{noi.eps,noi.pol}
it was  directly demonstrated  how  polarization actually
manifests itself as electrostatic fields in   nitride multilayers, due
to the polarization charges resulting from  polarization discontinuities
at heterointerfaces. This charge-polarization relation,
counterchecked in actual ab initio calculations,\cite{noi.pol} has been
exploited to calculate dielectric constants.\cite{noi.eps}

While piezoelectricity-related properties are largely standard,
spontaneous polarization is to some extent new in semiconductor
physics, to the  point that, so far, the practical importance of
spontaneous polarization in III-V nitrides nanostructures (multi
quantum wells, or MQWs, are the  focus of this paper) has
been largely overlooked. It is tantalizingly clear to us, however,
that these concepts may lead  to a direct and unambiguous measurement
of the spontaneous polarization in semiconductors, to the  recognition
of its importance in nitride-based nanostructures, and, hopefully, to
its exploitation in device applications. 
Our work has already spawned interpretations \cite{leroux} and
purposely planned  experiments\cite{cingo} in this direction;
 in the hope of accelerating
the process, in this paper we  show how to account for the effects
of spontaneous polarization  in MQWs, and discuss some prototypical
cases and their possible experimental realization.  
To support our arguments, we present simulations of typical 
AlGaN/GaN MQWs where spontaneous and piezoelectric polarizations 
are about equal.

Among the consequences of macroscopic polarization which we will
demonstrate in this paper let us mention the following:
{\it (a)}  the field caused by
the fixed polarization charge, superimposed on the compositional
confinement potential  of the MQW, red-shifts dramatically the transition
energies  and strongly suppresses interband transitions as the well 
thickness increases;   
{\it (b)} the effects of thermal carrier screening are negligible in
typical MQWs, although not in massive samples;  
{\it (c)}  a quasi-flat-band MQW profile can be approximately recovered
({\it i.e.} polarization fields can be screened) only in the presence  of
very high  free-carrier  densities, appreciably larger
 than those typical of semiconductor laser structures;
{\it (d)} even in the latter case,  transition
probabilities remain considerably smaller than  the ideal flat band
values, and this reduces quantum efficiency; 
{\it (e)} once an
appropriate screening density ({\it i.e.} the pumping power or injection current)
 has been
chosen to ensure that the recombination rate is sufficient, a
residual polarization fields is typically still present: this
 provides a means to intentionally red-shift transition energies by
changing well  thicknesses, without changing composition; 
{\it (f)} the
very  existence of distinct and separately controllable spontaneous
and piezoelectric polarization components allows to choose a
composition  such that they cancel each other out, leading to 
 flat-band conditions. Analogously, for a proper choice of
superlattice composition, piezoelectric  polarization can be made to vanish
and hence
a measure of spontaneous polarization can be accessed,
through {\it e.g.} the changes  in optical spectra.
It is clear that a fuller understanding of these points
will ultimately lead both to improvements in design and  operation of
real nitride devices, and to the direct measurement
of polarization, and a better knowledge thereof, in nitride semiconductors.

Before moving on, let us mention other recent contributions in this area.
Buongiorno Nardelli, Rapcewicz, and Bernholc,
using non--self-consistent effective-mass based 
perturbation theory in the small field limit, have predicted\cite{nardelli}
red shifts and transition probability suppression in InGaN quantum well;
Della Sala {\it et al.}, using self-consistent tight binding
calculations, have applied  \cite{ganapl} some of the ideas 
reported in this paper to InGaN/GaN quantum well lasers, explaining
several puzzling experimental features, among which the high thresholds
observed for GaN-based lasers, and several other aspects related to
self-consistent screening effects; Montecarlo simulations\cite{vogl} 
by Oberhuber, Vogl, and Zandler, employing the polarization calculated
  in  Ref.\onlinecite{noi.piezo}, have revealed
 a polarization-enhanced carrier density in the conduction channel
of   AlGaN/GaN HEMTs.
Takeuchi {\it et al.} have interpreted  \cite{takeuchi}  their own measurement 
of a quantum-confined Stark effect in InGaN/GaN superlattices as 
piezoelectricity-induced; this is, as it turns out, essentially correct in their
specific case involving InGaN, but not in general. Analogous results
have been reported in 
 Refs. \onlinecite{nak0} and \onlinecite{peng}, the latter including 
fairly detailed simulations,  but only accounting for piezoelectricity.
Finally, a detailed theoretical  exposition of effective-mass theory adapted
to deal with piezoelectric fields is given in Ref. \onlinecite{park},
including useful notation and basic formulas, and  some 
applications.  
 Experimental work will be mentioned later; here let us
 just quote the very recent evidence that  polarization-related 
effects on optical properties in selected AlGaN/GaN systems cannot be
 properly  interpreted if spontaneous polarization is neglected;
circumstantial evidence was obtained\cite{leroux} by Leroux  {\it et al.},
while a more carefully planned investigation, reaching firmer conclusions,
 has been carried out\cite{cingo} by Cingolani {\it et al}.

\section{Piezoelectric fields}
Piezoelectricity is a well known concept in semiconductor
physics. Binary compounds of strategic technological importance as
III-V  arsenides and phosphides can be forced to exhibit piezoelectric
polarization fields by imposing upon them a strain field.

Among others, applications of piezoelectric effects in semiconductor 
nanotechnology exist in the area of multi quantum-well (MQW) devices.
A thin semicondutor layer (active layer) is embedded in a
semiconductor matrix (cladding layers) having a different lattice
constant. 
If pseudomophic growth occurs, the active layer will be 
 strained and therefore subjected to a piezoelectric
polarization   field, which can be computed as
\be \Pvec^{\rm (pz)} =  \tensor{e} \cdot \vec{\epsilon}\, \ee
if the strain field $\vec{\epsilon}$ 
and the piezoelectric constants tensor  $\tensor{e}$ 
 are known.

In a finite system, the existence of a polarization field 
 implies the presence 
of electric fields. For the piezoelectric case, the magnitude of the 
latter  depends on strain, piezoelectric constants , and (crucially) on
  device geometry.
The structure of a typical III-V nitride-based superlattice or
MQW is  --C--A--C--A--C--A--C--
(A=active, C=cladding), where both the cladding layer  and the active
layer  are in general strained to comply with the substrate
in-plane lattice parameter. 
In such a structure,
the  electric fields in the A and C layers are
\be 
\Evec^{\rm (pz)}_{\rm A} =
 4\pi \ell_{\rm C} (\Pvec^{\rm (pz)}_{\rm C}
 - \Pvec^{\rm (pz)}_{\rm A})\, /(\ell_{\rm C}\,\veps_{\rm A} 
 + \ell_{\rm A}\,\veps_{\rm C}) \nonumber
\\
    \Evec^{\rm (pz)}_{\rm C} =
 4\pi \ell_{\rm A} (\Pvec^{\rm (pz)}_{\rm A}
 - \Pvec^{\rm (pz)}_{\rm C})\, /(\ell_{\rm C}\,\veps_{\rm A} 
 + \ell_{\rm A}\,\veps_{\rm C})
\label{eq.piezo}
\ee
where $\veps_{\rm A,C}$ are the dielectric constants  and $\ell_{\rm A,C}$ 
the thicknesses of  layers A and C. Thus, in general, an electric
field will be present whenever  $\Pvec_{\rm A} \neq \Pvec_{\rm
C}$. The above expressions are easily obtained \cite{nota3} by the
conditions that the electric displacement be conserved along the
growth axis, and by the boundary conditions that the potential energy
on the far right and left of the MQW structure are the
same. \cite{nota1} 

There are essentially three special cases of MQW structures  worth
mentioning: 
\begin{itemize}
\item[~~i)] 
active (cladding) layer lattice matched to the substrate:  $\Pvec_{\rm
A}= 0$ ($\Pvec_{\rm C}= 0$); \item[~ii)]
$\ell_{\rm A} = \ell_{\rm C}$, whence $\Evec_{\rm A} = -\Evec_{\rm C}$;
\item[iii)] $\ell_{\rm A}  \ll \ell_{\rm C}$, implying  $\Evec_{\rm C}
\simeq 0$, and  hence
\end{itemize}
\be \Evec^{\rm (pz)}_{\rm A} = 4\pi \Pvec^{\rm (pz)}_{\rm A} /\veps_{\rm A} 
.\ee
In the last case  we implicitly assumed the cladding layer
to be unstrained --
that is, its lattice constant to be relaxed to its equilibrium value 
because its thickness exceeds the critical value for pseudomorphic
growth  over the substrate. 
 $\Pvec^{\rm (pz)}$ may take any direction in general, but in 
normal practice its direction is parallel to the growth 
axis. \cite{nota}

To obtain piezoelectric  polarization effects in
zincblende semiconductor systems, lattice-mismatched epitaxial 
layers are purposely grown  along a polar axis, {\it e.g.} (111); the
in-plane strain propagates elastically onto the growth direction,
thereby generating $\Pvec^{\rm (pz)}$. In wurtzite nitrides, the
preferred growth direction is the polar (0001) [or (000$\overline{\rm
1}$)] axis, so that any non-accomodated in-plane mismatch automatically
generates a piezoelectric polarization along
 the growth axis (the sign depends on whether the epitaxial strain is 
compressive or tensile). We will be always be assuming this situation in
 the following. 

Usually, alloys are employed in the fabrication of MQWs. In that case,
one may estimate the piezoelectric polarization in the spirit of
Vegard's  law as, for a general strain imposed upon {\it e.g.} an
Al$_x$Ga$_{1-x}$N  alloy,
\be
    \Pvec^{\rm (pz)} 
= \left[ x ~\tensor{e}_{\rm AlN} + (1-x) ~\tensor{e}_{\rm
       GaN} \right]  	    ~\vec{\epsilon} \,(x)\, ,
\ee
This expression  contains terms linear as well as quadratic in $x$,
and similar relations hold for quaternary alloys. This piezoelectric
 term is only present in pseudomorphic strained growth, and will typically
tend to zero beyond the critical thickness  at which strain relaxation
sets in. Uncomfortable though it may be,\cite{zunger} the Vegard hypotesis is at
this point in time the only way to account for piezoelectric
(and spontaneous, see below) fields in alloys. As will be shown below,
indeed, the qualitative picture does not depend so much on the detailed
value of the polarization field as on their order of magnitude.

\section{Spontaneous  fields in MQWs}
New possibilities are opened by the use of III-V nitrides
(AlN,GaN,InN), that naturally cristallize in the wurtzite structure.
These materials are characterized by polarization properties that
differ dramatically from those of the standard III-V compounds
considered so far. From simple symmetry arguments,\cite{nye} it can
be shown that  wurtzite  semiconductors are characterized  by 
 a non-zero polarization in their equilibrium (unstrained) geometry,
named spontaneous polarization (or, occasionally, pyroelectric, with
reference to its change with temperature).\cite{nye2}
While the spontaneous polarization of ferroelectrics can  be  measured
via an hysteresis cycle, in a wurtzite this cannot be done, 
since no hysteresis can take place in that structure.
Indeed,  spontaneous polarization has never been measured directly 
in bulk  wurtzites so far.  III-V nitride MQWs offer  the opportunity to 
reveal its existence and to actually measure it. In turn, spontaneous
polarization can provide new degrees of freedom, in the form of
permanent {\it strain-independent}
 built-in electrostatic fields, to tailor transport and
optical characteristics of nitride nanostructures.  
 Its presence can {\it e.g.}  be exploited to cancel out the
piezoelectric fields produced in typical  strained
nitride structures, as discussed below.

Thanks to recent advances\cite{KS} in the modern theory of polarization
(a unified approach based on the Berry's phase concept),
it has become possible to compute easily and accurately from first
principles the values of the spontaneous polarization, besides
piezoelectric and dielectric  constants, in III-V
nitrides.\cite{noi.piezo,noi.eps} The results of the calculations
show that III-V nitrides  have important
polarization-related properties that set them apart from 
{standard} zincblende III-V semiconductors: 
\begin{itemize}
\item[~~i)]
huge piezoelectric constants (much larger than, and opposite in sign
to  those of all other III-V's);
\item[~ii)] existence of a spontaneous polarization  of the 
same order of magnitude as in ferroelectrics.
\end{itemize}
The latter is, we think, a most relevant property. Spontaneous
polarization implies  that even in heterostructure systems where
active and cladding layers are both lattice-matched to the substrate 
(so that no strain occurs, hence
no piezoelectricity), an electric field will nevertheless exist  due
to spontaneous polarization.
In addition, unlike piezoelectric polarization, spontaneous
polarization has a {\it fixed direction} in the crystal: in wurtzites
it is the (0001) axis, which is (as mentioned previously) the growth
direction of choice for nitrides epitaxy.  Therefore the field
resulting from spontaneous polarization will point along the growth  
direction, and this  {\it (a)} maximizes spontaneous
 polarization effects in these systems, and {\it  (b)} renders the
problem effectively one-dimensional.
In the simplest case of a fully unstrained (substrate lattice-matched)
MQW,  the electric  fields inside the layers are given, in analogy to
Eq. \ref{eq.piezo}, by
\be
\Evec^{\rm (sp)}_{\rm A} = 4\pi \ell_{\rm C} (\Pvec^{\rm (sp)}_{\rm C} 
- \Pvec^{\rm (sp)}_{\rm A})  
		       /(\ell_{\rm C}\veps_{\rm A} + \ell_{\rm
A}\veps_{\rm C}) \nonumber   
   \\
\Evec^{\rm (sp)}_{\rm C} = 4\pi \ell_{\rm A} (\Pvec^{\rm (sp)}_{\rm A} 
- \Pvec^{\rm (sp)}_{\rm C}) 
		       /(\ell_{\rm C}\veps_{\rm A} + \ell_{\rm A}\veps_{\rm C})
\label{eq.spont}
\ee
where the superscript ${\rm (sp)}$ stands for spontaneous;
 typical spontaneous  polarization values\cite{noi.piezo}
indicate that these fields are very large 
(up to several MV/cm).

\begin{figure}[h]
\epsfclipon
\epsfxsize=8cm
\centerline{\epsffile{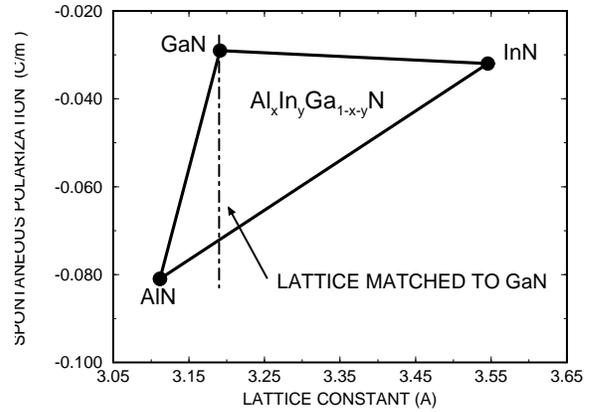}}
\caption{Spontaneous polarization in Al$_x$In$_y$Ga$_{1-x-y}$N
 alloys according to
a Vegard-like rule.}
\label{fig.vegard}
\end{figure}

 In actual applications  (for instance, to produce unstrained
MQWs)  alloys will have to be employed. The
values of the spontaneous polarization are accurately known only for
binary compounds.\cite{noi.piezo} In the absence of better estimates,
we assume as before that the spontaneous polarization in alloys can be
estimated  using a Vegard-like rule as
(for, {\it e.g.}, Al$_x$In$_y$Ga$_{1-x-y}$N)
$$    \Pvec^{\rm (sp)}(x,y) =
 x ~\Pvec^{\rm (sp)}_{\rm AlN} + y ~\Pvec^{\rm (sp)}_{\rm InN} 
                         + (1-x-y) ~\Pvec^{\rm (sp)}_{\rm GaN}\, .
$$
 In Figure \ref{fig.vegard} we report
the resulting  spontaneous polarization vs. lattice constant for the  
III-V nitrides,  with data   from Ref.\onlinecite{noi.piezo}.

\narrowtext
Figure \ref{fig.vegard} shows that for a given (substrate) lattice
constant, a  wide interval of spontaneous polarizations (hence of
spontaneous fields, according to Eq. \ref{eq.spont}) is accessible
varying   the alloy composition. In particular, consider a
GaN/Al$_x$In$_{y}$Ga$_{1-x-y}$N MQW,  where the composition is 
chosen so that the alloy be lattice matched to GaN, which we assume to
be also the substrate (or buffer) material (dashed-dotted line in
Figure \ref{fig.vegard}). Then, piezoelectric polarization vanishes,
but spontaneous polarization remains, and takes on values up to
$\sim$0.05 C/m$^2$. For a GaN quantum well with thick AlGaN
cladding layers, this means a theoretical electrostatic field of 
up to about 5 MV/cm.

\section{Fields in the general case}
 In general, of course, MQWs will be strained.
 Then,  for an arbitrary strain state, the electric fields 
in the A (or C) layers of the MQW are {\it the sum} of the 
piezoelectric and spontaneous
contributions: 
$$\Evec_{\rm A,C} = \Evec^{\rm (sp)}_{\rm A,C} + \Evec^{\rm (pz)}_{\rm
A,C},$$ 
where $\Evec^{\rm (pz)}$ is the old-fashioned piezoelectric field from
Eq.~\ref{eq.piezo}, and $\Evec^{\rm (sp)}$ is given by
Eq.~\ref{eq.spont}. It is important to stress  that 
$\Evec^{\rm (sp)}$ depends only on  material composition and not on
the strain state. Also, it is a key point to notice that although
both polarization contributions lay along the same direction
[the (0001) axis],   $\Pvec^{\rm (pz)}$  may have (due to its
strain dependence) the same or the
opposite sign with respect to the fixed  $\Pvec^{\rm (sp)}$ 
depending on the epitaxial relations.

\narrowtext
\begin{figure}[t]
\epsfysize=7cm
\centerline{\epsffile{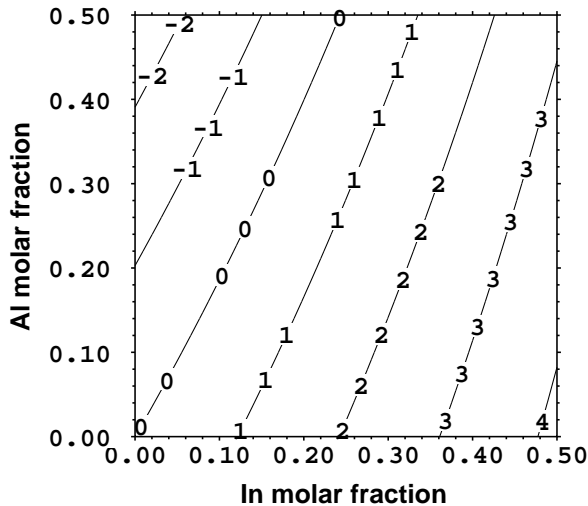}}
\caption{Total built-in electrostatic field in the active layer of a
  Al$_x$In$_y$Ga$_{1-x-y}$N/GaN 
MQW system (see text) vs. Al and In molar fraction.
Fields are in units of MV/cm, positive fields point 
 in the (0001) direction (Ga-face).}
\label{fig.efield}
\end{figure}

\begin{figure}[h]
\epsfysize=7cm\centerline{\epsffile{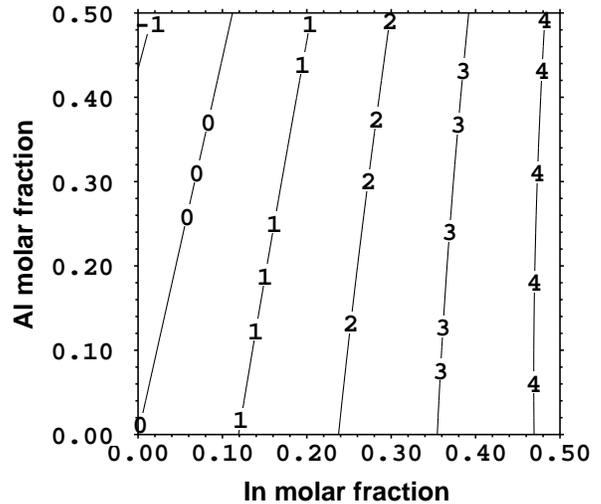}}
\caption{Piezoelectric component of the electrostatic fields
in the same MQW system as  in Figure\protect\ref{fig.efield} (see text).}
\label{fig.piezo}
\end{figure}

It is difficult to give a simple picture of the
electric field pattern in a general MQW system because of the many
degrees of freedom involved.  Here we consider an
Al$_{x}$Ga$_{y}$In$_{1-x-y}$N/GaN   MQW pseudomorphically grown over a
GaN  substrate, having active and cladding layers of the same thickness.
In such a case 
$$ \Evec^{\rm (sp)}_{\rm A} + \Evec^{\rm (pz)}_{\rm A} \equiv
\Evec_{\rm A} = - \Evec_{\rm C} \equiv -(\Evec^{\rm (sp)}_{\rm C} +
\Evec^{\rm (pz)}_{\rm C})\, . $$  
Note again, at his point, that the fields
(see Eqs. \ref{eq.piezo} and \ref{eq.spont}) are not related to just
the polarization of the material composing the specific layer, but a
combination of polarization {\it differences}, dielectric screening,
 and geometrical factors. \cite{nardelli2} We now consider the
field values in  the active layer: the total field $\Evec_{\rm A}$ 
is shown in Figure \ref{fig.efield} vs. Al and In molar  fraction; the
same is done for the piezoelectric  component  in
Figure \ref{fig.piezo}. In both cases the appropriate Vegard-like rules
have been used. 
\narrowtext
Comparison of these Figures cleary bears out the importance of
spontaneous polarization in determining the electric field.
 Several aspects  are worth pointing out. First, large electric
fields ($\sim$ 0.5--1 MV/cm) can be obtained already   for modest Al
and In concentrations. Second,  it is easy  to access compositions
such that Al$_x$In$_y$Ga$_{1-x-y}$N  is  lattice matched to GaN:
thereby, no piezoelectric fields exist, but  large, purely spontaneous
fields still do; specifically, this situation is realized for
compositions laying on the zero-piezoelectric-field line in
Figure \ref{fig.piezo}. On this locus of compositions,
spontaneous polarization is the only source of field and it can therefore 
be measured via the changes it induces in the MQW spectra.  
Third, it is possible to choose the material 
composition  in such a way that the active layers of a 
MQWs are free of electric fields. To achieve this situation the MQW must
be strained so that the piezoelectric and 
spontaneous polarizations cancel each other out; clearly, this is
realized for compositions laying on the zero-field line in
Figure \ref{fig.efield}. 
Of course the possibility of having a null field where desired
 is of capital importance in 
those devices  where electric fields in the active layer can not be
tolerated (other field screening mechanisms  are discussed
below).

 A noticeable feature of 
Figure \ref{fig.piezo} is that the piezoelectric component increases 
much faster with In content that with Al content, despite the larger
piezoelectric constants of the latter. The reason is, of course,
that strain builds up much more rapidly with In concentration.  
Along with the small difference in
spontaneous polarization between InN and GaN, this is the reason why
it is possible to interpret with reasonable accuracy polarization
effects
in InGaN/GaN structures on the basis of purely piezoelectric effects,
as done in Refs.\onlinecite{takeuchi,nak0,peng}.
On the other hand, it can be  seen that the {\it spontaneous}
component increases much more rapidly with Al content than with In
content, due to the widely different polarizations of AlN and
GaN. For AlGaN, piezoelectricity-based interpretations are bound to fail.

\section{Effects of polarization fields}
\label{screening}

We now come to the implications of polarization fields for devices
based on III-V nitrides.  In this Section we present a set of 
accurate self-consistent tight-binding  calculations for an isolated
AlGaN/GaN QW representing a system in which the
the  spontaneous-polarization contribution to the
total built-in  electrostatic field   
is as large as the piezoelectric term. To simulate realistically 
these nanostructures,  self-consistency is needed to describe field
screening by free carriers; the latter cannot physically cancel out
the polarization charge, which is fixed  and invariable, but may  
screen it out in part. In our calculations we therefore  solve
self-consistently the Poisson equation and  
the Schr\"odinger equation for a state-of-the-art
empirical tight binding Hamiltonian   for  realistic nanostructures.
\cite{aldo}
In the following, two cases are considered: {\it (a)} non-equibrium
carrier distribution (Subsec. A and B) related to photoexcitation or injection,  
where electron and hole quasi-Fermi levels are calculated  for a given
areal charge density ($n_{2D}$) in the quantum well (the {sheet density},
 related to  the injection current or optical pumping power); 
{\it (b)}  thermal equilibrium distribution 
(Subsec. C and D) where the Fermi level is calculated as a function of 
doping density by imposing charge neutrality conditions.\cite{aldo}  
We solve Poisson's equation,
\begin{equation} 
\frac{d}{dz}D =
\frac{d}{dz}\left(-\varepsilon\frac{d}{dz}V+P_T\right)=
e\left(p-n\right),
\label{eq:1}
\end{equation}
where the (position-dependent) quantities  $D$,   $\varepsilon$, and $V$,
 are respectively the displacement field, dielectric constant, and
potential. $P_T$ is the (position-dependent) total transverse
polarization. The effects of composition, polarization, and free
carrier screening are thus included in full. 
[Consistently with the aim of describing a {\it single} QW, we choose
 boundary conditions of  zero field at the ends of the simulation
region. This corresponds to the $\ell_{\rm C}\rightarrow \infty$ limit
in Eqs. \ref{eq.piezo} and \ref{eq.spont}.]

The potential thus obtained
is inserted in the Schr\"odinger equation,  which is solved
diagonalizing an empirical tight-binding $sp^3d^5s^*$
Hamiltonian. \cite{jancu} The procedure is iterated to self-consistency.
Further applications and details on the
technique can be found  elsewhere.  \cite{ganapl,aldo} 

Here we concentrate in particular on the
polarization-induced quantum-confined Stark effect 
(QCSE) in zero
external field, its control and
quenching, and its evolution with layer thickness.
 We first deal with the low free-carrier densities regime:
thereby the QCSE  manifests
itself as a strong red shift of the interband transition energy, with a
concurrent suppression of the transition probability, both of these
features getting stronger as the well thickness  increases. This is
the regime that applies to low-power operation or
 photoluminescence experiments. 

Next we discuss how the QCSE can be modified, and eventually (almost)
quenched, by providing  the QW with a sufficiently high free-carrier
density. In this regime, as the free carrier density increases,  the
transition energy is progressively blue-shifted back towards its flat
band value, and the transition probability suppression is largely
removed. The needed free-carrier density depends on the polarization
field, and not surprisingly it is found to be  very
substantial. Typical values of the sheet density range in the
10$^{13}$ cm$^{-2}$, as opposed to typical values of
10$^{12}$  cm$^{-2}$ needed to obtain lasing in GaAs-like materials.

\subsection{QCSE at low power}

\begin{figure}[ht]
\epsfclipon
\epsfysize=8cm
\centerline{\epsffile{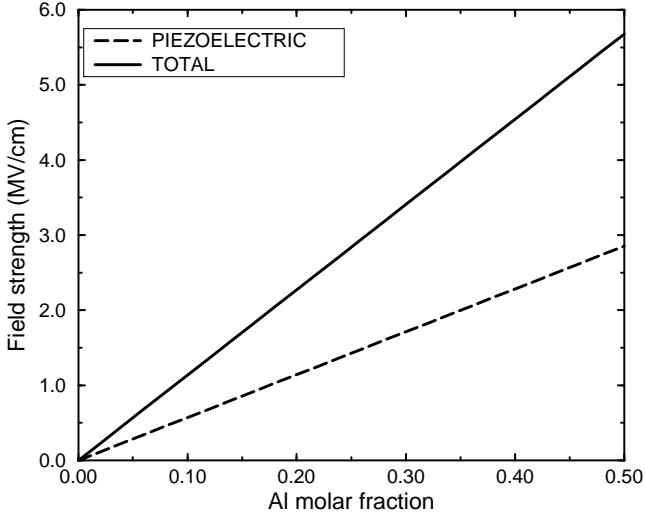}}
\caption{Total field and its piezoelectric component in
the GaN/Al$_x$Ga$_{1-x}$N QW discussed in the text.}
\label{fig.piezo.2}
\end{figure}

The prototypical system we consider is an isolated  GaN 
quantum well cladded between  Al$_{x}$Ga$_{1-x}$N
barriers. In Figure \ref{fig.piezo.2}, we display the
total field {\bf E}$_{\rm A}$ in the (isolated) active well, and its
piezoelectric component as a function of the
Al molar fraction $x$. The spontaneous component is the difference of
the two, and therefore approximately equal  to
 the piezoelectric one.\cite{nota2}

The value we pick for our simulations is  $x$=0.2, a
reasonable compromise between the conflicting needs for not-too-large fields,
 sufficient  confinement,
 and technologically achievable composition.
In this case the valence offset is  $\Delta E_v=0.064$ eV.
The total field in the QW of --2.26 MV/cm, and the spontaneous and
piezoelectric components are --1.14 MV/cm and --1.12 MV/cm
respectively. The minus  signs indicates that the field points
in the  (000$\overline{1}$) direction. The bare polarization charge
at the interface is  proportional to the change in polarization
 across the interfaces, and it
amounts to $\sim$1.28$\times$10$^{13}$ cm$^{-2}$. The {\it field}
value mentioned above  results from this charge as screened by the
dielectric response of the QW (the field change at the interface is
thus related to a smaller, or screened, effective interface charge\cite{noi.pol}). 

We  performed a series of calculations for different well widths, 
where the
electron and  hole confined states have been populated ({\it i.e.} pairs
have been created) with a density of $\sim$
10$^{11}$ cm$^{-2}$ to simulate a low-power optical excitation.
We find  that this density has only a  very marginal effect:
indeed, the potential is perfectly  linear, {\it i.e.} the electrostatic field
remains uniform, over the whole QW.
The square-to-triangular change in the potential shape causes a small
blue shift 
of both the electron and hole confined states (referred to the flat 
well bottom), but the linear potential
given by the field causes a much larger relative red shift for any reasonable
thickness. Also, since the thermal carrier density fluctuations are 
negligible at microscopic thicknesses and room temperature (see below,
and Ref.\onlinecite{mermin}), one expects  the QW band edge profile
to  remain linear as a function of thickness, at least for the low
excitation powers  typical of photoluminescence spectroscopy.     
 
\begin{figure}
\epsfclipon
\epsfxsize=7.5cm
\centerline{\epsffile{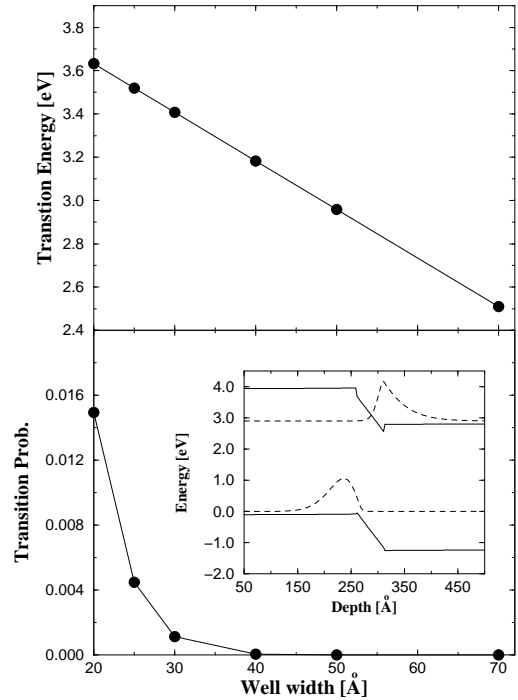}}
\caption{Transition energy red shift and suppression of transition probability
vs well thickness. In the inset self-consistent band edge (solid), 
electron and hole envelope functions of the TB wavefunctions (dashed) for 
a 50 \AA ~thick QW.
}
\label{fig.low.2}
\end{figure}

In Figure \ref{fig.low.2}
we show the TB result for the lowest interband transition energy and the 
corresponding  transition probability 
({\it i.e.} the  squared overlap 
of the highest
hole level and the lowest electron level envelope wavefunctions~\cite{aldo}) 
as a function of QW
thickness. 
Both the Stark red shift and the strong suppression 
of the transition probability are evident. This was to be expected 
from the potential shape and the reduced overlap of hole and electron 
states, displayed in the inset of Figure \ref{fig.low.2}.

It is worth noting that the localization of the hole envelope function in
 the well region is rather weak, because 
the large effective field blue-shifts the hole bound state energy 
close to  the valence barrier edge. This will generally be the case for
 low-$x$ AlGaN wells, due to  the small valence confinement 
energy.\cite{noi.pol} Indeed, even the
 conduction confinement is  small
 on the scale of the  fields-induced
 potential drop, and the electron bound state also tends to
 have the character of a  resonance for small $x$ ({\it i.e.} small confinement).

We conclude that  in the absence
of excitation and at normal operation temperatures, or 
at low optical excitation power, macroscopic
polarization fields  cause QW's to be highly inefficient in 
emitting light, and the emission energy to be 
considerably different than the gap of the material plus
the  confinement energies. 

Comparison with experiment is tricky since most attempts to measure
these effects are polluted by inappropriate (at least for the purpose
of revealing polarization effects)  choices of the experimental
 geometry. For instance, measurements have been done\cite{leroux}
 in a {\it series} of quantum wells of different thicknesses ranging
between 10 and 50 \AA.\cite{leroux2} 
 In any case, the general experimental features
\cite{leroux,nak0,takeuchi,hangl} are in full agreement with the notion that the
transitions are red-shifted essentially linearly with increasing well
thickness, and that screening at low free carrier densities is
irrelevant in this class of systems. This is not quite true any more
for thick layers, as will be discussed in Sec.\ref{sec.mass}.   

\subsection{QCSE quenching at high excitation power}

If  carriers are generated optically, one can envisage  that a
sufficiently high excitation power could possibly produce the carrier
density needed to screen the polarization field. We now calculate the
properties of the QW as a function of the  free-carrier areal density,
to check if the red shift and the  transition probability suppression
can be removed in a physically accessible range of such density.     

\begin{figure}
\epsfclipon\epsfysize=9cm\centerline{\epsffile{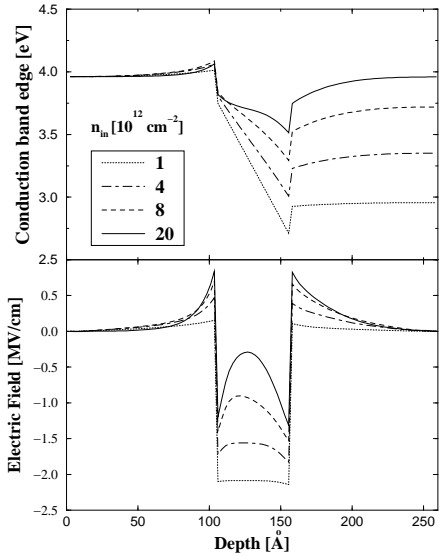}}
\caption{Self-consistent band edge (a) and electric field (b) 
for various sheet densities ($\sim$ excitation level) in a 50 \AA ~thick
QW.}
\label{fig.high.1}
\end{figure}

\begin{figure}
\epsfclipon
\epsfxsize=7cm
\epsfysize=7.5cm\centerline{\epsffile{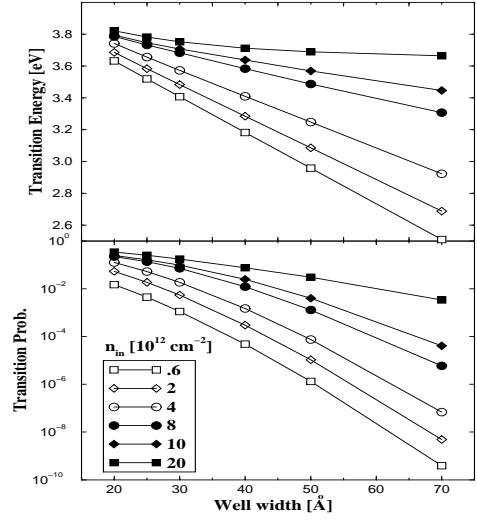}}
\caption{Removal of red shift and recovery of interband transition probability
upon high excitation.}
\label{fig.high.2}
\end{figure}

We repeat the self-consistent procedure  increasing progressively the
free charge density in the QW from  10$^{12}$ up to 2 $\times$
10$^{13}$ cm$^{-2}$.  We see in Figure \ref{fig.high.1} that,
albeit  at the cost of a large increase of the QW free-carrier
density, the field does get progressively screened.
As can be
seen in Figure \ref{fig.high.2}, at fixed thickness the  red shift
decreases as the carrier density increases,
and it tends to become thickness-independent at the highest densities.
The transition probability is also increased by several orders of magnitude.
However, the field is not screened abruptly but 
dies off gradually, with an effective screening length of about 20
{\AA} for the largest density used here (of course, this is a token of
the larger spatial extension of the screening charge as compared to
the polarization charge \cite{noi.pol}): herefore, holes and electron
remain spatially separated to a large extent even at high carrier
densities, and the trasition probability never quite  goes back to
unity. This is  likely to be one of the reasons  for the
relatively low quantum efficiency observed in typical nitride MQW
devices. For the same reasons, the  transition energy never quite goes back
 to the flat-band value (gap  plus confinement energy). Note in
passing that because of strain,   in these calculations  $E_g^{\rm
GaN} = 3.71$ eV, almost 10 \% larger than the equilibrium value.

The screening density  of order 2 $\times 10^{13}$ cm$^{-2}$ needed
to partially screen out the field   corresponds to an 
estimated \cite{niwa} optical pumping power of
about 10 to 20 kW/cm$^2$ per well. This figure 
  agrees nicely with the unusually  high
pumping powers\cite{leroux,nak0,hangl,domen} needed to obtain the laser effect in nitride
structures. The explanation is simply that much of the free
 charge being generated actually goes into screening the polarization    
field.  On the other hand, our result prove that the optically activated 
lasing conditions can indeed be realized in practice, although with
high pumping powers, so that there seems to be  no need to invoke
quantum dot  formation\cite{nardelli2} or other exotic effects to
explain lasing in nitride structures. On the other hand, the same
phenomenon  explains the high current threshold observed for
electrically driven GaN  based lasers.\cite{nakamura:2,fang:1,yeo:1} 

QCSE quenching phenomena similar to those just described have been
observed \cite{takeuchi} by Takeuchi {\it et al.}  in InGaN/Gan MQWs,
with estimated fields in the 1 MV/cm range. The red shift and
optical inefficiency were in fact removed , although only in part and in 
a transient fashion, by sufficiently high excitation powers.
The order of magnitude of the values reported in Ref.\onlinecite{takeuchi} is
$\sim$200 kW/cm$^2$ for 5 to 10 MQW periods, {\it i.e.} 20 to 40 kW/cm$^2$
per well, in qualitative agreement with our estimate above.

 One important
remark at this point is  that,  depending on the   excitation power,  
 the MQW will adsorb radiation at  many different
transition energies ranging from that of the built-in-field--biased
well (low power limit) to the quasi--flat-band well (high power limit)
-- that is, the MQW acts a multistable switch. It is indeed fortunate
that the typical fields in these structures are such that one can 
physically access  the various possible regimes.

Another noticeable effect is that at a properly chosen value of the
sheet density ({\it i.e.} of the excitation power) one can obtain at the
same time a reasonable transition probability {\it and} a red-shifted
energy by just  increasing the well thickness. This is very useful since
the transition wavelength can be shifted to a different color without
changing alloy composition, but only the well thickness. For instance
(see Figure \ref{fig.high.2}),
changing the well thickness from 20 to 30 {\AA} at a sheet density of
4$\times$10$^{12}$ cm$^{-2}$,  one obtains an energy red shift of
0.1 eV at the cost of a
 loss of a factor 10 in recombination rate, which may still be acceptable
depending on the application. Red-shifting the transition energy in
this fashion may avoid the need to add {\it e.g.} some In in the QW
composition. Of course, 
{\it blue}-shifting by thickness reduction will 
 increase the transition probability.

\subsection{Self-screening of fields in massive samples}
\label{sec.mass}

Free charge produced by  high excitation  screens polarization fields
fairly efficiently over the quite short distances typical 
 in nanostructures, basically because the spatial extension of the
screening charge 
is comparable to the size of the system. How about extended samples,
especially if not subjected to  illumination, {\it i.e.} having only
intrinsic free   carriers ?   It is indeed the case that no
macroscopic 
 fields exist in ``infinitely large'' samples even in the absence of
high densities  of (say) photogenerated carriers. The simple  reason
is that the  intrinsic carrier fluctuations in an undoped semiconductor rise
 exponentially as a function of deviations of the chemical potential
 from the  mid-gap value.\cite{mermin} In polarized  nitrides, such
 deviations occur due to the built-in fields. As the sample thickness
increases,  the potential drop grows linearly. When the drop is
smaller than the  gap, the field is uniform:
$|\Evec|=4\pi\Pvec/\varepsilon_0$.   When the drop approaches  the gap
value, {\it i.e.} for thicknesses approaching  $d_c=E_{\rm
gap}/|\Evec|$, the Fermi level nears the band edges:  consequently,
large amounts of  holes and electrons   are generated on the opposite
sides of the sample. These intrinsic carriers
  screen partially the polarization charges,
preventing the gap from closing.  The total potential drop is thus
pinned at the gap value for all  thicknesses $d>d_c$ -- that is, the
effective gap decreases down to zero, but not  
below. For $d>d_c$,  the field will decrease  as
 $$|\Evec|=E_{\rm gap}/d.$$  For this picture to hold, the spatial
extension of the  screening charge at the sample surface  must be
comparable with that of  the polarization charge (a few {\AA}
at most \cite{noi.pol}) and much smaller that the sample size. This  
 will cause  the field  inside the sample to remain uniform, since the net
 effect of screening will be to change the effective polarization
 charge. Indeed, this assumption turns out to be verified in
 practice on direct inspection, as we discuss below.

Clearly,   the above mechanism will strongly influence QW's of 
 thicknesses equal to, or larger than, the critical value $d_c$.
For the system we are considering here, with a built-in field
of --2.26 MV/cm, the critical value is $d_c \sim$ 165 {\AA}.
To confirm our picture, we  simulated  QW's with the same composition and
geometry considered in Subsec. B, and thicknesses below and above 
 $d_c$, to mimic the crossover
from a ``microscopic'' to a ``macroscopic'' sample.  
In this case, we need to describe very extended bulk regions
on  the left and right of the QW, in order to account for the large
 screening length. Thus, we have made use of a classical Thomas-Fermi
 model, where the charge densities are calculated with
 Fermi-Dirac statistics of a classical  system rather than  by solving
 the Schr\"odinger equation in the TB basis. 
This allows to consider devices with a spatial extension of several
hundreds of microns. Effective masses in this calculation were fitted
  to accurately reproduce the self-consistent 
TB results.
 
\begin{figure}
\epsfclipon
\epsfysize=7cm\centerline{\epsffile{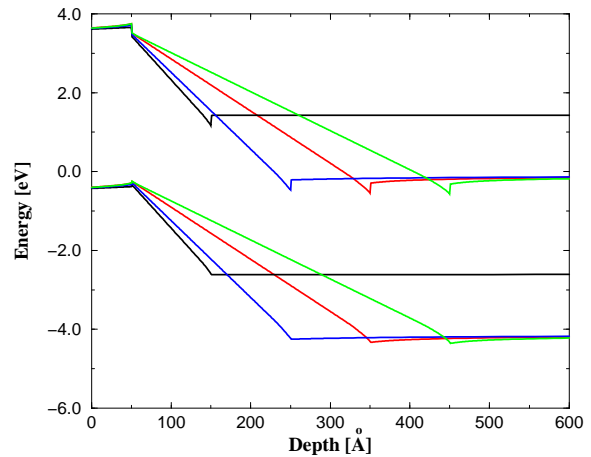}}
\caption{Field self-screening in a thick layer near and
above the critical thickness.}
\label{fig.mass.1}
\end{figure}

The resulting 
self-consistent potential is shown in Figure \ref{fig.mass.1} for 
well thicknesses of 100, 200, 300, and 400 {\AA}.
 A first point to note is that the field remains uniform for 
all well thicknesses. The field value
equals the polarization field 
for the smallest thickness (smaller than $d_c$); for the thicker
wells, the field (while  remaining uniform) indeed decreases
progressively as $\propto 1/d$. 

Photoluminescence experiments are not expected to be able to reveal
this effect (which should cause a saturation of the red shift 
as function of thickness) in very thick QW's, since the effective
recombination rate rapidly  becomes vanishingly small.
Experiments aiming to reveal this effect should be
designed considering   our result, that a very thick layer 
is effectively subjected to a uniform electrostatic field
$E_{\rm gap}$/$d$. In an unstrained GaN QW, for
$d>d_c$ (the latter being typically of order 100-200 {\AA} or
so depending on the polarization) the field is   $\sim$3.4 V/$d$, {\it i.e.}
 $\sim$70 kV/cm for $d=$0.5 $\mu$m. This is presumably sufficient to cause  
observable bulk-like effects such as  shifts in response functions or field
 effects on impurities.

A similar ``self-screening'' behavior has been  revealed
 indirectly   in devices comprising sufficiently thick layers. In
Ref. \onlinecite{yu} a 300 {\AA} thick  Al$_{0.15}$Ga$_{\rm 
0.85}$N layer was grown on a very thick GaN substrate, and topped with
a Schottky contact. The predicted field in the AlGaN layer is 1.4
MV/cm, which would cause 
 a  potential drop 
of 4.2 eV across the layer.  The  maximum reasonable potential
drop dictated by 
Schottky barriers, conduction offset,  and Fermi level  is about  1
eV, so it must be the case that the polarization charge gets largely
screened by electrons from the GaN layer,  forming a high-density
two-dimensional electron gas (2DEG) at the heterointerface; this, by
the way, causes an enhanced conductivity in the active channel. CV
depth profiling indeed reveals a 2DEG at the interface.\cite{yu}
An equivalent, more formal description is that the  field
 would force the metal-determined Fermi level to some 3 eV
above the conduction band of GaN, thus attracting towards the
interface  an enormous carrier density, which  screen out (part of)
 the field.   Note in passing that in Ref. \onlinecite{yu} only
 piezoelectric polarization  was considered, which leads to an
 underestimation of the 2DEG density, since the piezoelectric
 contribution is actually  about one third of the total interface
 charge.  Similar considerations apply to other similar
 experiments.\cite{bykhovski1}  The recent device simulation of
 Ref. \onlinecite{vogl} has corrected this point, including in part
 the spontaneous-polarization interface charges.   

\subsection{Suppressing QCSE by doping}

We have seen in the previous Sections that polarization fields can be
screened to a reasonable extent by generation of  free charge of
{\it both} kinds in the QW  upon {\it e.g.} optical excitation.
Qualitative problems with this screening mechanism are
that {\it (a)}  it is transient, since it disappears when photoexcitation
or current injection are removed, and that
{\it (b)} in purely electronic ({\it i.e.} non-optoelectronic) devices,
it is unlikely that the high densities needed can be reached in normal
operating conditions. Besides, the  current is not constant
in time, so that the well shape also changes in time.

\begin{figure}
\epsfclipon
\epsfysize=7cm\epsfxsize=9cm\centerline{\epsffile{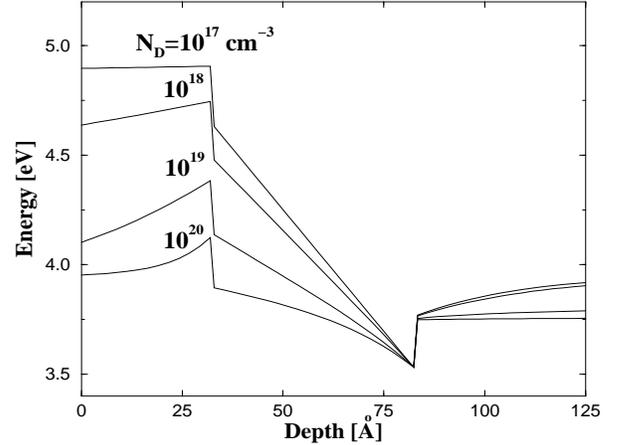}}
\caption{Conduction band edge of a remotely-doped GaN/AlGaN quantum
 well.}
\label{fig.dope}
\end{figure}

It is natural to presume that the same effects can be achieved in a
permanent fashion using extrinsic carriers from dopants. 
The idea is to provide the well
with carriers which would screen the polarization charge, excepts that
now  the electrons are released into the QW from the doped barriers,
and not injected or photogenerated. 
Of course, this effect is not
transient as the others discussed previously.  
The problem
is, how high must the doping density be to achieve the same level of
screening as in a high optical excitation regime.
We simulated 
a 50 {\AA} thick Al$_{0.2}$Ga$_{0.8}$N/GaN single QW,
where the barriers have been doped $n$- type in the range from 
10$^{17}$ to 10$^{20}$ cm$^{-3}$ and the donor ionization energy\cite{aln} 
 have been set to 10 meV. In this 
simulation, we used again the selfconsistent TB approach. 
The resulting conduction band profile is displayed in 
Figure \ref{fig.dope} for the various doping densities. 
%
The polarization field raises the conduction band on the left side 
over the Fermi energy,  and 
in order for the barrier conduction band to reach the Fermi level on
the far left,
the electrons are transferred from the left-side barrier into the QW,
leaving behind a large depletion layer.
As a consequence of the electron flow into the QW, the polarization field
start to get significantly screened at doping densities  
above $\sim$ 10$^{19}$cm$^{-3}$. 
The existence of a depletion layer causes 
a large band bending in the left-side barrier, while the bending is absent 
in the left half of the well. This is  quite different to the case 
of the photoexcited well, where the bending on the left side of the 
well was due to hole accumulation  near the interface (see Figure 6).
This explains why the field remains nearly uniform in the left half 
of the well for all of the simulations performed.
On the right side of the well, only a small bending due to the electron 
accumulation is present.
Indeed electron localization is quite weak in these systems, since 
the confinement potential is 
small as compared to the field-induced drop, and electrons
tend to spill over to the right-side barrier. This is
likely to be case in all nitride systems in this composition range.

From these results, we conclude that doping can indeed
be used to screen polarization fields. While  
it is not obvious that the needed doping level can always be reached in
practice, it is likely  
that a combination of doping and 
current injection or photoexcitation
will generally succed in quenching polarization fields
in the range of MV/cm, thus allowing for 
recovery of quasi flat band conditions. 
Fields in  InGaN/GaN systems will be
generally smaller than  
those in AlGaN/GaN systems for typical compositions in use today,
and will therefore be more easily amenable to treatment  by 
the above technique.
This  procedure has in fact been adopted in experiment by
Nakamura's group,\cite{nak0} which  reported that a doping level of 10$^{19}$
cm$^{-3}$  is sufficient to quench the QCSE to a large extent.
Indeed, in their In$_{0.15}$Ga$_{0.85}$N/GaN MQWs 
the unscreened field is  $\sim$1.2 MV/cm, {\it i.e.} approximately a half of
the one we considered here. This field will be more easily screened by remote 
doping, in qualitative  agreement with our findings. 


\section{Summary and acknowledgements}
In conclusion, we have discussed how macroscopic (and in particular,
spontaneous) polarization 
plays an important  role in nitride-based MQWs by producing large
built-in electric fields. Contrary to zincblende  semiconductors, in
III-V nitrides--based devices the spontaneous polarization is an
unavoidable source of large electric fields even in lattice-matched
(unstrained) systems.   The existence of these fields may also be used
as additional degree of freedom in device design: for instance, for
an appropriate choice of alloy composition, spontaneous and
piezoelectric fields may be caused to cancel out, thus freeing the
structures from built-in fields.
We have also discussed the different regimes of  free carrier
screening, effected by doping or optical excitation, showing that 
fields can be screened  only in the presence of high free carrier
densities,
which leads to unusually high lasing thresholds for undoped QW's.
Of course, our results about the effects on the electronic structure
apply qualitatively to any kind of polarization field, thus in
particular also to piezo-generated  ones.

VF and FB acknowledge special support from the PAISS program of
 INFM. VF's stay at the Walter Schottky Institut  was supported by the
 Alexander von Humboldt-Stiftung. FDS, ADC and PL acknowledge support from 
Network Ultrafast and 40\% MURST.


\end{multicols}

\begin{thebibliography}{99}

\bibitem{noi.piezo}
F. Bernardini,  V. Fiorentini, and D. Vanderbilt, Phys. Rev. B {\bf 56},
R10024 (1997).

 \bibitem{noi.eps}
F. Bernardini,  V. Fiorentini, and D. Vanderbilt,
Phys. Rev. Lett. {\bf 79}, 3958 (1997); F. Bernardini and V. Fiorentini, 
Phys. Rev B {\bf 58} 15292 (1998).

\bibitem{noi.pol}
F. Bernardini and  V. Fiorentini,
Phys. Rev. B {\bf 57}, R9427 (1998).


\bibitem{leroux}
M. Leroux, N. Grandjean, M. La\"ugt, J. Massies,
B. Gil, P. Lefebvre, and P. Bigenwald,
 Phys. Rev. B {\bf 58}, 13371 (1999).

\bibitem{cingo}
R. Cingolani, M. Lomascolo, G. Col\`{\i}, F. Della Sala, A. 
Di Carlo, and P. Lugli,  to be published.

\bibitem{nardelli}
M. Buongiorno Nardelli, K. Rapcewicz, and J. Bernholc,
Appl. Phys. Lett. {\bf 71}, 3135 (1997).

\bibitem{ganapl}
F. Della Sala, 
A. Di Carlo, P. Lugli, F. Bernardini, V. Fiorentini,
R. Scholz, and J.-M. Jancu, Appl. Phys. Lett. {\bf 74}, 2002 (1999).

\bibitem{vogl}
R. Oberhuber, G. Zandler, and P. Vogl, Appl. Phys. Lett. {\bf 73}, 818 (1998).

\bibitem{takeuchi} T. Takeuchi, S. Sota, M. Katsuragawa, M. Komori,
H. Takeuchi, H. Amano and I. Akasaki, Jpn. J. Appl. Phys. {\bf 36}, L382.

\bibitem{nak0}
T. Deguchi, A. Shikanai, K. Torii, T. Sota, S. Chichibu,
and S. Nakamura,  Appl. Phys. Lett. {\bf 72}, 3329 (1998);
S. Chichibu {\it et al.}, Appl. Phys. Lett. {\bf 73}, 2006 (1998).

\bibitem{peng}
L.-H. Peng, C.-W. Huang, and L.-H. Lou,
Appl. Phys. Lett. {\bf 74}, 795 (1999);

\bibitem{park}
S.-H. Park and S.-L. Chuang, Phys. Rev. B {\bf 59}, 4725 (1999) and
references therein.

\bibitem{nota3}
From dielectric displacement conservation (see Ref.\onlinecite{noi.eps})
 one gets $4\pi (P_{\rm C}-P_{\rm A}) =
\varepsilon_{\rm A} E_{\rm A} -\varepsilon_{\rm C} E_{\rm
C}$. Periodicity implies  $\ell_{\rm A} E_{\rm A} + \ell_{\rm C}
E_{\rm C} =0$. Solving for  one of the fields, 
Eqs.\ref{eq.piezo} are immediately obtained.

\bibitem{nota1}
This assumption is valid in the limit in which the electric field  
across the whole structure is negligible 
with respect to the internal fields induced by polarization.

\bibitem{nota}
Note that for normal devices layered along the nominal 
growth direction, the electric field is always parallel to the growth 
direction,  even if the polarization is not.

\bibitem{zunger}
Vegard-like estimates of the polarization should of course be
considered with caution, as it is   established (see {\it e.g.}  P. Ernst,
C. Geng, M. Burkard, F. Scholz,  and H. Schweizer, in {\it The Physics
of Semiconductors},   M. Scheffler and R. Zimmermann eds. (World
Scientific, 1996),  p. 469;  S. Froyen, A. Zunger, and A. Mascarenhas,
Appl. Phys.   Lett. {\bf 68}, 2852 (1996))  that ordering in (cubic)
III-V solid  solutions can produce  spontaneous polarization, an
effect  not unexpected also in the XN's. Even in the random solution,
short-range order in the form of  bond alternation  may  alter the
local electronic structure, hence the  polarization. We are currently
investigating this problem.

\bibitem{nye}
J. F. Nye, {\it Physical properties of crystals} (Oxford UP, 1985).

\bibitem{nye2} 
One possible way to see it is as the
zero-strain limit of the piezoelectric 
polarization.

\bibitem{KS}
R. D. King-Smith and D. Vanderbilt, Phys. Rev. B {\bf 47}, 1651 (1992).
R. Resta, Rev. Mod. Phys. {\bf 66}, 899 (1994).


\bibitem{nardelli2}
This contradicts  some of the assumptions and conclusions of
Ref.\onlinecite{nardelli}.

\bibitem{aldo}
A. Di Carlo {\it et al.}, Solid State Communications {\bf 98}, 803
(1996);  A. Di Carlo, MRS Proc. {\bf 491} 389 (1998).

\bibitem{jancu}
J.-M. Jancu, R. Scholz, F. Beltram, and F. Bassani,
Phys. Rev. B {\bf 57}, 6493 (1998);


\bibitem{nota2}
Since we use the {\it calculated} lattice constants of III-V nitrides
to determine strains,
the piezoelectric component may  be overestimated
due to the slight underestimation of $a_{\rm AlN}$.

\bibitem{mermin}
N. W. Aschroft and N. D. Mermin, {\it Solid State Physics},
(Holt-Saunders Japan, Tokio 1981), p.576 and p.611.


\bibitem{leroux2}
In Ref.\onlinecite{leroux} the red shift was found to be compatible 
with a field of 450 kV/cm, to be compared with the predicted values of
piezoelectric field, 120 kV/cm, and total field, 640 kV/cm.
The obvious source of uncertainty in the polarization values extracted 
from this  experiment are of course the
intricate and not uniquely defined
 boundary conditions produced by the awkward choice of geometry.
We note further that the agreement with the total field would improve
in the case of epitaxial relaxation.


\bibitem{hangl}
J. S. Im, H. Kollmer, J. Off, A. Sohmer, F. Scholz, and A. Hangleiter,
Phys. Rev. B {\bf 57}, R9435 (1997).


\bibitem{niwa} A. Niwa, T. Ohtoshi, and T. Kuroda, 
Jpn. J. Appl. Phys., {\bf 36}, L771 (1997); I. Nomura, K. Kishino,
A. Kikuchi, Solid-State Electron.  {\bf 41}, 283 (1997);
 R. J. Radtke, U. Waghmare, H. Ehrenreich, and C. H. Grein, 
App. Phys. Lett. {\bf 73}, 2087 (1998)


\bibitem{domen} K. Domen, A. Kuramata, and T. Tanahashi, 
App. Phys. Lett., {\bf 72}, 1359 (1998).

\bibitem{nakamura:2}
 S Nakamura {\it et al.},
App. Phys. Lett. {\bf 72}, 2014 (1998); {\it ibid.} {\bf 73}, 832
(1998).

\bibitem{fang:1} W. Fang, and S. L. Chuang, App. Phys. Lett.
{\bf 67}, 751 (1995).

\bibitem{yeo:1} Y. C. Yeo, T. C.Chong, M. F. Li, and W.J. Fan,
J. App. Phys. {\bf 84}, 1813 (1998).


\bibitem{yu}
E. T. Yu, G. J. Sullivan, P.M. Asbeck, C. D. Wang, D. Quiao, and
S. S. Lau,  Appl. Phys. Lett. {\bf 71}, 2794 (1997).

\bibitem{bykhovski1} 
R. Gaska, J.W. Yang, A.D. Bykhovski, M.S. Shur,
V.V. Kaminskii and S. Soloviov,  Appl. Phys. Lett. {\bf 71}, 3817
(1997).


\bibitem{aln}
Doping of Al$_{x}$Ga$_{1-x}$N  becomes rapidly inefficient
above x$\sim$0.45 (see {\it e.g.} 
A. Fara, F. Bernardini, and V. Fiorentini, J. Appl. Phys. {\bf 85}, 2001
 (1999), and references therein),
but it is viable at the low $x$ of interest  here.
\end{thebibliography}
\end{document}